\documentclass[%
twocolumn, 
showpacs,preprintnumbers,
 amsmath,amssymb,
 aps,
prb,
citeautoscript,
floatfix, 
]{revtex4} 

\usepackage{graphicx}
\usepackage{dcolumn}
\usepackage{bm}

\showboxdepth=\maxdimen
\showboxbreadth=\maxdimen

\begin{document}

\preprint{}

\title{Selective nonresonant excitation of vibrational modes in suspended graphene via vibron-plasmon interaction}

\author{Axel M. Eriksson}
\email{marer@chalmers.se}

\author{Leonid Y. Gorelik}
\affiliation{%
Department of Applied Physics, Chalmers University of Technology,\\ Kemig\aa{}rden 1,
412 96 G\"{o}teborg, Sweden}%



\date{\today}

\begin{abstract}
We theoretically study a doped graphene ribbon suspended over a trench  and subject to an ac-electrical field polarized perpendicularly to the graphene plane. In such a  system, the external ac-field is coupled to the  relatively slow mechanical vibrations via plasmonic oscillations in the isolated graphene sheet. We show that the electrical field generates an effective pumping of the mechanical modes. It is demonstrated that in the case of underdamped plasma oscillation, a peculiar kind of geometrical resonance of the mechanical and plasma oscillations appear. Namely the efficiency of pumping significantly increases when the wave number of the mechanical mode is in close agreement with the wave number of the plasma waves. The intensity of the pumping increases with the wave number of the mode. This phenomenon allows selective actuation of different mechanical modes although the driving field is homogeneous.
\end{abstract}

\pacs{73.22.Pr, 81.07.Oj,46.70.De}
\maketitle
Since the first graphene sample was isolated and studied experimentally, the experimental and theoretical work on the 2D material has grown tremendously due to its many extraordinary properties\cite{Novoselov,Terrones2010}. The high mobility, low mass and mechanical strength of graphene makes it well suited as the basis of nanoelectromechanical resonators. The frequency tunability and high quality factor of  graphene based resonators make them promising for e.g., mass sensing\cite{Chen2009} and filtering applications\cite{Xu2010}. To actuate the nanomechanical resonators, different principal schemas are utilized. First of all, mechanical oscillations can be initiated  by applying an electrical field at resonance frequency with the mechanical vibration\cite{Chen2013,Meerwaldt2012,Lassagne2009}. Another method which is utilized to control mechanical motion  exploits the radiation pressure induced by an electromagnetic field in an optomechanical cavity \cite{Barton2012,Heikkila2014}. In this case the external frequency, at which the system is driven, is nonresonant with the relatively low mechanical frequency. Nonresonant excitation of the mechanical vibrations can also be achieved by integration of the mechanical resonator
in an electrical LC-circuit\cite{Brown2007}. In both cases, the force acting on the mechanical subsystem is determined by the detuning of the external frequency and the resonance frequency of the  cavity or external LC-circuit. The resonance frequency depends on the mechanical displacement which induces an electro-mechanical time-delayed backaction. The backaction generates an effective pumping (or damping) of the mechanical vibrations. Therefore, it is possible both to excite and cool the resonator. These phenomena have been demonstrated for many systems\cite{Metzger2004,Kippenberg2008} and for graphene based resonators in particular\cite{Barton2012}. Recently it was shown that similar effects can be achieved by integrating the resonator into an RC-circuit\cite{Eriksson2015}. In this description, the actuation mechanism was due to the time-delayed overdamped charge response rather than coupling via a resonant high-frequency mode.

In this article, we show that nonresonant excitation of mechanical vibrational modes can be achieved also for an $isolated$ graphene membrane via 
its internal charge dynamics.   We will demonstrate that the nonresonant actuation mechanism presented here enables selective actuate of different mechanical modes, even antisymmetric ones. The intensity of actuation increases with mode number in contrast to the optomemechanical and electrical pumping mechanisms mentioned above where predominantly actuation of the fundamental mode takes place.     

\section{Model}
\begin{figure}\centering
\includegraphics[scale=0.4]{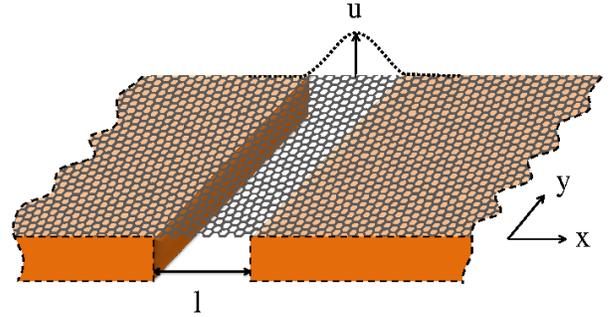}
\caption{ An isolated graphene membrane is suspended over a cavity inside a wave-guide. The suspended part of the membrane is free to perform vertical vibrations. A electrostatic ac-field is applied polarized perpendicularly to the graphene sheet.}
\label{fig1}
\end{figure}

The sketch of the system under consideration is presented in Fig. \ref{fig1}, and comprises
the isolated graphene ribbon suspended over a trench with width $l$. The system is placed in a wave guide. Electromagnetic waves, with wave length  much larger than $l$ travel inside the wave guide along the $x$ axis. The field is assumed to be homogeneous in the trench and screening effects from the wave guide is neglected. The wave is polarized perpendicularly to the flat membrane and induces an electrical field along the membrane only when it is deflected from its flat position. The induced field generate electronic charge waves in the graphene sheet. Simultaneously, the electrical field  exerts a force on the suspended part when it is charged and provides a feedback coupling between the electronic and the mechanical subsystems.

To analyse this feedback we model the free vibrating part of the ribbon as an elastic membrane.
For simplicity we consider the membrane to be infinite in the y-direction. In this limit, we assume that the membrane deflection $U(x,y,t)=U(x,t)$, charge density $\varrho(x,y,t)=\varrho(x,t)$ and current density $j_{x}(x,y,t)=j(x,t)$ are uniform along the trench. We disregard the geometric nonlinearity of the graphene membrane since it does not affect the nonresonant phenomenon discussed in this paper and can be neglected at small amplitude of oscillation $U\ll l$. 

Under these assumptions the dynamical equations for the flexural out-of-plane modes become
\begin{equation}
\label{fullDyn}
 \left(\frac{\partial^2}{\partial t^2}+\gamma \frac{\partial}{\partial t}-\frac{T_0}{\varrho_m}\frac{\partial^2}{\partial x^2}+\frac{\kappa}{\rho_m}\frac{\partial^4}{\partial x^4}\right)U(x,t)=\frac{\varrho(x,t)}{\rho_m } E(t)
\end{equation}
with intrinsic mechanical damping $\gamma$, built-in tensile stress $T_0$, bending rigidity of graphene $\kappa$, electrical field in the wave guide $E(t)=E_0\cos(\Omega t)$ and 2D-mass density of graphene $\rho_m$. The corresponding boundary conditions of the clamping are $U(x,t)=0$ and $U'(x,t)=0$ at $x=\pm l/2$.
The membrane deflection $U(x,t)$ can be presented as a superposition of the vibrational eigenmodes
 \begin{eqnarray}\label{U)}
 U(x,t)=\sum_{n=1}^{\infty}u_{n}(t)f_{n}(\xi)   \\
 \left( -\frac{1}{\pi^2}\frac{\partial^2}{\partial \xi^2}+b^2\frac{\partial^4}{\partial \xi^4}\right)f_{n}(\xi)=K^{2}_{n}f_{n}(\xi) \nonumber \end{eqnarray}
with dimensionless spatial coordinate $\xi=x/l$, stretching-bending ratio $b^2=\kappa/\pi^2 l^2 T_0$ and $f_{n}(\xi)$ is the normalized spatial profile of the flexural  eiqenmodes.

To describe the charge  dynamics of the electronic subsystem we will use a simple hydrodynamic  approach \cite{Svintsov2012,Jablan2013}. We will consider monopolar electronic plasma where the Fermi  energy $E_{F}$ is much greater than temperature and $\hbar\Omega$.
Within this approach the charge evolution is described by
\begin{eqnarray}
\label{gidr}
 \frac{\partial}{\partial t}\varrho(x,t)= - \frac{\partial}{\partial x}j(x,t),\\
\label{gidr2}
 \frac{\partial}{\partial t}j(x,t)+\nu j(x,t) = \frac{1}{\mathcal{L}}E(t,x),
\end{eqnarray}
where $1/\mathcal{L}=e^{2}E_{F}/\hbar^{2}\pi$ and $\nu$  is the scattering frequency. The electrical field along the ribbon consists of one external and one internal contribution
\begin{equation}\label{E}
  E(t,x)= E(t) \frac{\partial}{\partial x}U(x,t)+\frac{1}{2\pi\epsilon_0}\mathcal{P}\int \frac{\varrho({x_{1}},t)}{x-x_{1}}\textrm{d} x_{1}.
\end{equation}
The first term in Eq. (\ref{E}) describes the external electrical field induced along the membrane when it deflects from its flat position. The second term describes the internal non-local electrostatic field due to charge redistribution.

The time-scales of the system is obtained by consider a typical experimental situation where we take $l\approx 10\ \mu$m, $E_{F}\approx 1$ meV, end  $T_{0}\approx 0.1$N/m which gives $b\sim 10^{-4}$. 
Under such conditions the characteristic mechanical frequency $\omega_{M}\approx 100$ MHz and the characteristic plasma frequency $\omega_{p}\sim\nu\approx 1$ THz are well separated. 
Further, we consider high-frequency external driving $\Omega\sim\omega_{p}\gg\omega_{M}$. The strength of the electromechanical coupling generated by the external field is  characterized by
the coupling frequency $\omega_{E}=E_0\sqrt{\epsilon_0/2\pi l\rho_m}$. We will consider low amplitude external field so that $\omega_{E}$ is the smallest frequency $\omega_{E}<<\omega_{M}<<\Omega$.

\section{Effective mechanical dynamics}
To get the coupled dynamics for the amplitudes $u_{n}(t)$ and charge density $\varrho(\xi,t)$ we combine Eqs.(\ref{gidr}) and (\ref{gidr2}) and obtain
\begin{eqnarray}
\label{fullsetMech}
\ddot{u}_{n}(t)+\gamma u_{n}(t) + \omega^{2}_{M}K_{n}^{2}u(t)
 = \frac{E(t)}{\rho_m} \langle f_{n}(\xi), \varrho(\xi,t)\rangle,\\
\ddot{\varrho}(\xi,t)+ \nu\dot{\varrho}(\xi,t)-\frac{\omega_{p}^{2}}{\pi^{2}}\frac{\partial}{\partial \xi}\mathcal{P}\int \frac{\varrho(\xi_{1},t)d\xi_{1}}{\xi-\xi_{1}}=\nonumber\\
-  \frac{E(t)}{\mathcal{L}l^2}\sum_{n}u_{n}(t)\frac{\partial^2}{\partial\xi^2}f_{n}(\xi),
\label{fullsetChar}
\end{eqnarray}
where $\langle f_{n}(\xi), \varrho(\xi,t)\rangle$ denotes projection of the charge distribution on the spatial mode function $f_n$.
The characteristic mechanical and plasma frequencies $\omega_{M}$  and  $\omega_{p}$ are defined in table \ref{tabfreq}.
\begin{table}
\caption{Dispersion relation for plasma and mechanical vibrations. The continuous wave number $q$ corresponds to a plasma wave length $\lambda=2l/|q|$. }\
\begin{center}
\begin{tabular}{c|c}
 Electronic & Mechanical\\ \hline
$\omega(k)=\omega_p\sqrt{|q|}$ & $\omega_n=\omega_M K_n$\\
$\omega_p=\frac{e}{\hbar}\sqrt{\frac{E_F}{2\epsilon_0 l}}$ & $\omega_M=\frac{\pi}{ l}\sqrt{\frac{T_0}{\rho_m}}$
\end{tabular}
\end{center}
\label{tabfreq}
\end{table}

The electrostatic forces acting on the vibrational modes can be expressed by substituting an integral expression for the charge density described by Eq. (\ref{fullsetChar}) in the right hand side of Eq. (\ref{fullsetMech}). The forces $\mathcal{F}_{n}(t)=E(t) \langle f_{n}(\xi), \varrho(\xi,t)\rangle/\rho_m$ can then be formulated as 
\begin{eqnarray}\label{Force}
\mathcal{F}_{n}&(t)=&\\
\nonumber \omega_{E}^{2}&\sum_{m}&\int^{t}_{-\infty}\textrm{d}t_{1} G_{nm}(t-t_{1})\cos(\Omega(t-t_{1}))u_{m}(t_{1}) +\nonumber \\
\nonumber\omega_{E}^{2}&Re& \left[ e^{2i\Omega t}\sum_{m}\int^{t}_{-\infty}\textrm{d}t_{1} G_{nm}(t-t_{1})e^{i2\Omega(t-t_{1})}u_{m}(t_{1})\right],
\end{eqnarray}
with
\begin{equation}\label{GrF}
G_{nm}(t)=\omega_{p}^{2}\int \frac{2e^{-\nu t/2}\sin\left(t\sqrt{\omega_{p}^{2}|q|-\nu^{2}}/2\right)}{\sqrt{\omega_{p}^{2}|q|-\nu^{2}}}w_{nm}(q)\textrm{d}q.
\end{equation}
where $w_{nm}(q)=-(\pi q)^{2}\langle f_{n}(\xi), e^{i\pi\xi q}\rangle \langle e^{-i\pi\xi q},f_{m}(\xi)\rangle$.

The electrostatic forces on the form Eq. (\ref{Force}) introduce linear feedback on the mechanical motion. The feedback on mode $n$ is direct back to itself via $G_{nn}(t)$ but the feedback also couples different modes via $G_{nm}(t)$ $n\neq m$. We want to note that the subsets of odd and even modes do not couple. However, since the coupling strength $\epsilon=\omega_E/\omega_M$ is assumed to be small we disregard the coupling between modes since it will affect the mechanics only to fourth order in $\epsilon$. The system of equations Eq. (\ref{fullsetMech}) then decouples to independent single mode oscillators.

The dynamics is further simplified since we consider the high-frequency regime of the driving frequency $\Omega\sim\omega_{p}>>\omega_{M},\omega_{E}$. As we will see later, under such conditions only modes with $K_{n}\simeq (\Omega/\omega_{p})^2$ play an important role in the membrane dynamics. We seek the time evolution of the amplitudes $u_{n}(t)$ in the form of perturbation series
\begin{equation}\label{u}
  u_n(t)=\sum_{m=0}^{\infty}\varepsilon^{2m} e^{i2m\Omega t}u_{n,m}(t)
\end{equation}
here $\varepsilon =\omega_{M}/\Omega <<1$ and $u_{n,m}(t)$ are slow on the time scale $\Omega^{-1}$. The second term in the right hand side of Eq. (\ref{Force}) gives corrections of the order $\epsilon^2\varepsilon^2$. We will neglect corrections of this order of smallness and take $u_n(t)\approx u_{n,0}(t)$. In  these approximations, the dynamics of mode $n$ is governed by
\begin{eqnarray}\label{AmEv}
\ddot{u}_n(t)+\gamma\dot{u}_n(t)+\omega_M^{2}K_n^{2}u_n(t)= \\
 \omega_{E}^{2}\int_{-\infty}^{t}\textrm{d}t_{1}\ G_{nn}(t-t_{1})\cos(\Omega(t-t_1))u_n(t_1).
\end{eqnarray}

 The dispersion relation which characterises the time evolution of the $n$:th mode can be obtained by the Ansatz $u_n(t)=\exp(i\omega_{n} t)$. Substituting this form in Eq. (\ref{AmEv}) we  obtain
\begin{eqnarray}\label{DR}
  -\omega^{2}_{n} +i\gamma\omega_{n} + \omega^{2}_{M}K_{n}^{2}=\omega_{E}^{2}G_{nn}(\omega_{n};\Omega)\\
 G_{nn}(\omega;\Omega)=-\int_{-\infty}^0 \textrm{d}t \exp(-i\omega t)G_{nn}(t)\cos(\Omega t)
\end{eqnarray}
Solving Eq. (\ref{DR}) we arrive at the following approximation (with an accuracy of $\epsilon^{2}$) for the complex frequencies $\omega_{n}$
\begin{eqnarray}
\omega_{n}&=&
   \omega_{M}\left[K_{n}+ \frac{\epsilon^{2}}{K_{n}}\Lambda_{n}(\tilde{\Omega})+\frac{i}{2}\left(\frac{1}{Q} + \frac{\epsilon^{2}\tilde{\omega}_{M}}{K_n}\eta_{n}(\tilde{\Omega})\right)\right] \nonumber \\
\Lambda_{n}(\tilde{\Omega})&=&\int\Lambda(\tilde{\Omega},q)w_{nn}(q)dq,\,\nonumber \\
\eta_{n}(\tilde\Omega)&=& \int\eta(\tilde\Omega,q) w_{nn}(q)dq.
\end{eqnarray}
here $Q=\omega_{M}/\gamma$ is the characteristic Q-factor and we have introduced the dimensionless frequencies $\tilde{\omega}_{M}=\omega_{M}/\omega_{p}$ and $\tilde\Omega=\Omega/\omega_p$. The functions $\Lambda(\tilde{\Omega},q)$ and $\eta(\tilde\Omega,q)$ are given by
\begin{eqnarray}
\Lambda(\tilde{\Omega},q)&=\frac{1}{2}\frac{\tilde{\Omega}^2-|q|}{\left(|q|-\tilde{\Omega}^2\right)^{2}+\tilde{\nu}^{2}\tilde\Omega^2} \\
\eta(\tilde{\Omega},q)&= -\tilde\nu\frac{ \left(|q|+\tilde{\Omega}^2\right)^{2}-\tilde{\Omega}^2\left(4\tilde{\Omega}^2+\tilde{\nu}^{2}\right)}{\left(\left(|q|-\tilde{\Omega}^2\right)^{2}+\tilde{\nu}^{2}\tilde\Omega^2\right)^{2}}
\end{eqnarray}
and $w_{nn}(q)=(\pi q)^{2}|\langle e^{i\pi\xi q},f_{n}(\xi)\rangle|^{2}$. To calculate $w_{n}(q)$  we neglect the bending rigidity of the graphene in comparison with the clamping tension since $b<<1$.  Under this condition we can take $f_{n}(\xi)=\sin(\pi n\xi+\pi n/2)\theta(1/2-|\xi|)/\sqrt{2}$ with $K_{n}= n$. These  forms for the mode shapes give
\begin{equation}\label{w}
  w_{nn}(q)=\frac{2q^{2}n^{2}\sin^{2}(\pi (|q|+n)/2)}{(q^{2}-n^{2})^{4}}.
\end{equation}

\section{Selective mode actuation}
\begin{figure}\centering
\includegraphics[scale=0.7]{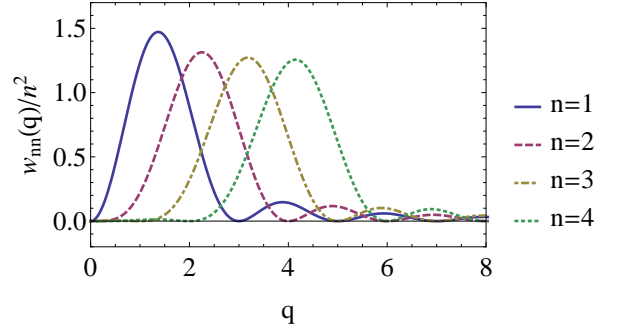}
\caption{Spatial geometrical resonance between the charge oscillations and the mechanical mode functions occur in the vicinity of $q=K_n=n$. Wave number $K_n=n$ corresponds to a mechanical mode with wave length $\lambda=2l/n$.}
\label{wnn}
\end{figure}

 The shift of the complex frequencies occure at strong coupling between the mechanical modes and the charge waves. The charge waves are mainly generated from the regions close to the clamping since the gradient of the electrical field along the membrane is biggest in these areas. The coupling is strong when the wave length of the generated charge waves is in close agreement with the mechanical wave length. This can be seen in Fig. \ref{wnn} since the functions $w_{n}(q)$ has a sharp maximum in the vicinity of $|q|\sim n$.
 Therefore we have a peculiar kind of spatial geometrical resonance but a nonresonant phenomenon in the time domain.
Simultaneously, at small plasma damping $\tilde{\nu}<<1$, the functions $\eta(\tilde{\Omega},q)$ and $\Lambda(\tilde{\Omega},q)$ dramatically increase when $|q|=\tilde\Omega^{2}$.

\begin{figure}\centering
\includegraphics[scale=0.67]{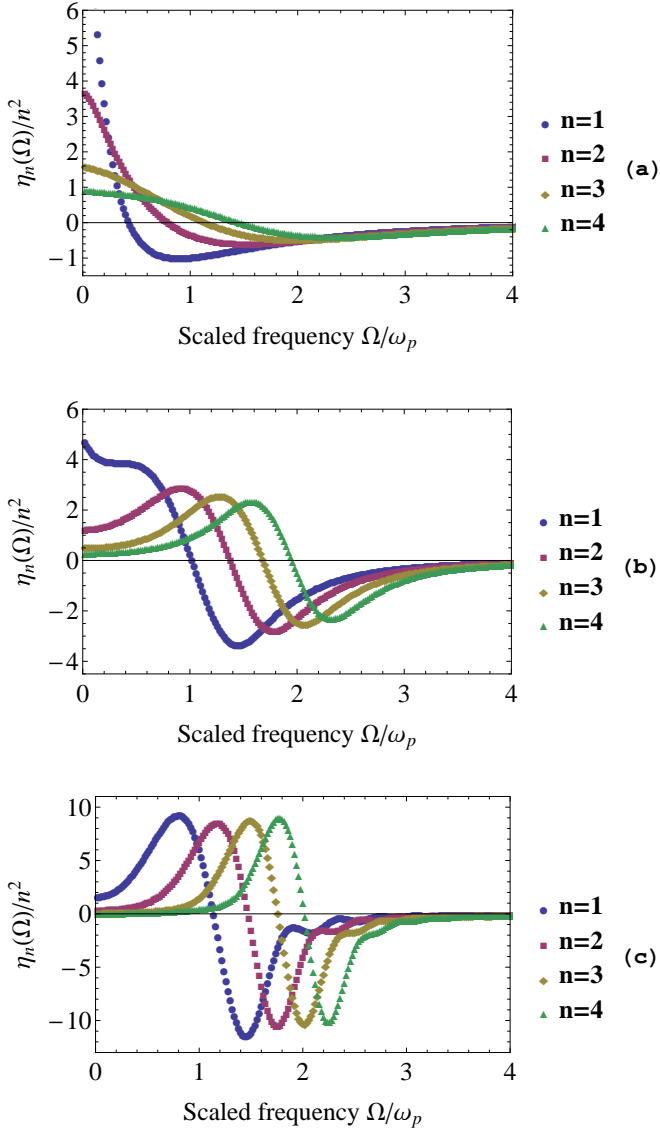}
\caption{The damping coefficient for damping ratio $\tilde\nu$ equal 3, 1 and 1/3 in (a), (b) and (c), respectively. A region of negative damping coefficient is present at driving frequencies $\Omega>\omega_p$. The mechanical modes can selectively be driven if $\tilde\nu<1$ due to the geometric resonance of plasma and mechanical oscillations.}
\label{fig3}
\end{figure}

The shift of the mechanical damping is qualitatively described by the normalized damping coefficient $\eta_{n}(\tilde\Omega)/n^{2}$, Fig. \ref{fig3}.
The damping coefficient of the $n$th mechanical mode $\eta_{n}(\Omega/\omega_{p})$   becomes negative at $\Omega = \Omega_{n}^{c}(\tilde{\nu})\sim \omega_{p}$ and reaches its minima $\eta_{n}^{min}(\tilde{\nu})$ at the minima frequency $\Omega_{n}^{min}(\tilde{\nu})$.
If the plasma oscillation is overdamped Fig. \ref{fig3}a, the characteristic width of the minima is much greater than the distance between the minima frequencies $\Omega_{n}^{min}$, while $\eta_{n}^{min}(3)\approx - n^{2}$. In the underdamped situation Fig. \ref{fig3}c, the distances between the minima frequencies become greater than the width of the minima and pumping strength $\eta_{n}^{min}(1/3)\approx -10 n^{2}$. It should be particulary emphasised that the distance between minima frequency as well as minima widths are three order of magnitude greater than, and independent of, the characteristic mechanical frequency. Because of this, the phenomenon is nonresonant in the time domain.

A vibrational mode will become mechanically actuated if the effective pumping generated by the high-frequency external field overcomes the intrinsic mechanical damping of the mode. 
To actuate a mode the driving frequency has to be in the region where the electromechanical cpupling gives negative damping and the amplitude of the external field has to exceed a critical value $E_n^c(\Omega)$. Above the critical value the mechanical vibration is unstable and will be saturated by nonlinear effects. The field strength needed to achieve this can be estimated by using $\rho_m\approx 0.7 $ mg/m$^2$,  assuming the quality factor to be $Q=10^5$ and damping ratio $\nu/\omega_p=1/3$. This gives an estimate of the critical field strength for the fundamental mode $E_1^c\approx30$ V/$\mu$m, at the optimal driving frequency $\Omega_1^c$.

Selective actuation of vibrational modes is possible when the overlap of the minima peaks is small, Fig. \ref{fig3}c. This possibility is remarkable since the applied electrical field is homogeneous. It is interesting to note that in contrast to optomechanical excitation where only symmetric modes are actuated and the strength of pumping decreases with mode number, whereas in our system also antisymmetric modes can be actuated and the strength of pumping increases with mode number.

From the above analysis it follows that pronounced selective nonresonant excitation of the mechanical modes is achievable for  $\tilde{\nu}\ll1$. However, there are natural restrictions for $\tilde{\nu}$ and $n$ which come from the applicability of the hydrodynamic description of the charge dynamics used in this paper. The hydrodynamic equations are not valid in the ballistic regime of electronic propagation. To analyze the range of parameters where our approach is valid, it is convenient to introduce the effective electron mean free path $l_{sc}=v_{F}/\nu$.
We then have the following expression for the damping ratio $\tilde{\nu}=\nu/\omega_{p}=0.27 l/(l_{sc}\sqrt{k_{F}l})\simeq 0.07l/l_{sc}$. The hydrodynamic approach fails when the electron mean free path exceeds characteristic space variations in the system $l_{sc}>l/n$. Therefore, the model formulated here is not valid for $n>14\tilde{\nu}$.

\section{Conclusions}
To conclude, we have shown that  the internal charge dynamics in a suspended isolated graphene sheet can be utilized to selectively actuate vibrational modes by a nonresonant homogeneous external field. The phenomenon occurs when the external field induces plasma oscillations with a wave length comparable to the wave length of the spatial profile of the vibrational mode. Different modes can then be selectively driven via this geometrical resonance, if the plasma oscillations are underdamped. 
\begin{acknowledgments}
The authors thank the Swedish Research Council (VR) for funding our research.
\end{acknowledgments}

\bibliography{article}

\end{document}